\documentclass[12pt]{iopart}

\usepackage{graphicx,dcolumn,bm,amssymb,amsmath,latexsym,footnote}
\usepackage{mathrsfs}
\usepackage[utf8]{inputenc}
\usepackage[unicode,breaklinks=true]{hyperref}
\allowdisplaybreaks[1]

\newtheorem{lemma}{Lemma}

\newtheorem{theorem}{Theorem}

\newcommand{\nonu}{\nonumber}

\newcommand{\noi}{\noindent}

\newcommand{\bp}{\begin{picture}}
\newcommand{\ep}{\end{picture}}
\newcommand{\bc}{\begin{center}}
\newcommand{\ec}{\end{center}}
\newcommand{\be}{\begin{equation}}
\newcommand{\ee}{\end{equation}}
\newcommand{\beal}{\begin{align}}
\newcommand{\eeal}{\end{align}}
\newcommand{\bea}{\begin{eqnarray}}
\newcommand{\eea}{\end{eqnarray}}

\newcommand{\Pint}

\begin{document}

\title[Multipole moments in axistationary spacetimes and the $N$-soliton solution]{The relations between the multipole moments in axistationary electrovacuum spacetimes and the $N$-soliton solution}

\author{Etevaldo dos Santos Costa Filho$^{1,}$, Angelo Guimar\~{a}es $^{2,}$\, and I. Cabrera-Munguia$^{3,}$}

\address{$^1$ Instituto de F\'isica de S\~{a}o Carlos-USP, 13566-970 S\~{a}o Carlos, S\~{a}o Paulo, Brazil}
\address{$^2$ Instituto de Ci$\hat{e}$ncias Matem\'aticas e Computac\~{a}o de S\~{a}o Carlos-USP, 13566-970 S\~{a}o Carlos, S\~{a}o Paulo, Brazil}
\address{$^3$ Departamento de F\'isica y Matem\'aticas, Universidad Aut\'onoma de Ciudad Ju\'arez, 32310 Ciudad Ju\'arez, Chihuahua, M\'exico}
\eads{\mailto{etevaldocostaf@gmail.com}, \mailto{g.angelo@usp.br},\mailto{icabreramunguia@gmail.com}}

\vspace{10pt}
\begin{indented}
\item[]April 2021
\end{indented}

\begin{abstract}
A general formulation of the relativistic multipole moments in axistationary electrovac spacetimes is introduced in a closed analytical form. We give a complete description of how the Ernst potentials on the symmetry axis can be completely characterized by these arbitrary multipole moments. Some concrete applications are also developed.
\end{abstract}

\section{Introduction}

Multipole moments for static asymptotic flat spacetimes have been first defined by Geroch \cite{GerochI,GerochII}. Later on,  Hansen \cite{Hansen} extended this result to the stationary case. Other authors have also interpreted and proposed novel approaches to compute the multipole moments  \cite{Thorne,Simon-Beig,Quevedo,Simon,FHP,HP,Sotiriou,fodor2020calculation}.

It is well-known that a spherically symmetric black hole solution like the Reissner-Nordstr\"{o}m one \cite{Reissner,Nordstrom} can be entirely characterized by the mass and electric charge, being the two monopole moments of the gravitational and electric fields, respectively. The same statement can be announced for Kerr's black hole \cite{Kerr} where now the first two multipole moments; the mass and angular momentum, define the source perfectly in stationary spacetimes. In both scenarios, the black hole solutions fulfill the no-hair conjecture \cite{Israel,Carter,Robinson}.

On the other hand, the exterior gravitational field of a neutron star cannot be explicitly comprehended unless multipole moments of higher orders are considered, such as the moment of inertia and quadrupole moment, in addition to its mass and radius, since, for instance, they might have some deformation in the mass distribution \cite{PhysRevD.91.103003}. Here, the multipole moments can even play a whole in trying to understand the equation of state for a star \cite{Laarakkers_1999}. In this context, the study of multipole moments is really useful for a better understanding of the physical measurable quantities of any type of sources in General Relativity with an astrophysical meaning; for instance, black holes or neutron stars \cite{Berti15,BertiStergioulas}, and other interesting physical features related to them like geodesics, shadows, and lensing effects or quasinormal modes, among others. 

 As a matter of fact, Fintan D. Ryan \cite{Ryan95,Ryan97} provided in 1995 a practical application for the multipole moments by extracting physical information from the gravitational radiation emitted by an object orbiting around a central body (see \cite{Pappas15b} for example in scalar-tensor theories). Actually, the recent detection of gravitational waves produced by the coalescence of binary black hole mergers \cite{LIGO}, suggest that the measure of the spin-induced quadrupole moment in compact binary mergers permits a clear distinction between binary black hole systems and binary systems containing another type of exotic compact objects \cite{Krishnendu}. In this regard, it is quite clear that the multipole structure of astrophysical objects is per se of great physical and mathematical relevance, and it deserves to be taken into account.

Several years ago, in 1996, a complete description for multipole moments in stationary vacuum systems had been accomplished by Hernández-Pastora, J.L. \cite{pastora}. He was able to work out a mapping among the $4N$-parameter exact stationary solution with the $4N$ arbitrary Geroch-Hansen multipole moments \cite{GerochI,Hansen}. This $4N$ exact solution is the vacuum specialization of the $N$-soliton electrovac solution derived in \cite{RMJ} via the Sibgatullin method  \cite{Sibgatullin,Manko1993}; recalling that Sibgatullin's method provides an exact stationary solution in the entire spacetime once it is established any specific form of the Ernst potentials  \cite{Ernst}  on the symmetry axis (the axis data). In 1998, Manko and Ruiz published this result again in a section of a relatively famous paper \cite{MankoRuiz}, but now lacking of proof and containing fewer details. The main idea of the thesis \cite{pastora} and the paper \cite{MankoRuiz} is quite clear; the parameters composing the axis data can be related to arbitrary multipole moments and vice-versa; therefore, the whole spacetime can be completely represented with a whole physical meaning. Naturally, one might expect to extend this result to the electrovacuum scenario, which to our knowledge, has not been investigated yet (or just ignored due to its complexity).

Moreover, a natural demand due to the increase of numerical models fitting astrophysical observations is to construct analytical solutions \cite{1995ApJ...444..306S,PhysRevLett.108.231104} in order to bring more qualitative and quantitative understanding in what has been measured—making more evident the need to directly relate the physical quantities to the mathematical parameters that appear in the generating techniques.

The present paper aims to introduce a concise general formulation for electrovac spacetimes in terms of a multipole moment structure, extending the earlier results provided in \cite{pastora}. Ernst and Hauser have shown that the Ernst potentials fall into the class of elliptic differential equations, implying that they satisfy a quite interesting and useful property: the behavior of the Ernst potentials, $\mathcal{E}$ and $\Phi$, on the symmetry axis is sufficient to perform an analytic continuation of them
to the whole space. Meaning that, in order for us to fully connect a solution of the Einstein-Maxwell equations in axistationary spacetimes with the multipole moments, we only need to consider the relationship between the Ernst potentials and the multipole moments on the symmetry axis. The outline of our paper is the following. In Sec. \ref{Nsol} the N-soliton solution for electrovac spacetimes \cite{RMJ} as well as some basic backgrounds are first explained. Afterward, the path provided by Hernández-Pastora, J.L. \cite{pastora} is revisited in Sec. \ref{multipolevacuum} with the main purpose of entering the reader to the electrovacuum case that is described in Sec. \ref{multipoleelectrovacuum}.  In Sec. \ref{examples} the multipole structure is developed for the particular case 2-soliton solution, where two results previously considered in the literature \cite{TomiSato} are taken into account to test the validity of our result and we also discuss how to use the $N$-soliton to construct approximated solutions possessing arbitrary multipole moments and use the monopole-dipole source as an example.  

\section{The Extended N-soliton solution}\label{Nsol}

The use of solitonics techniques for constructing exact solutions of the Einstein equations was first introduced by  Belinski and Zakharov in 1978 \cite{Belinsky:1971nt,Belinsky:1979mh} by means of their formulation of the inverse scattering method (ISM).  In particular, the study of stationary axisymmetric spacetimes possesses an enormous physical interest because they can describe, in an idealized way, for instance, the exterior region of black holes, neutron stars, and accretion flow. 

The stationary and axisymmetric spacetimes, which admits $G_2$ as isometry group, possess a set of completely integrable equations \cite{stephani_kramer_maccallum_hoenselaers_herlt_2003,martini1985geometric}. Several authors have introduced generating techniques based on these symmetries, and most of them, constructed employing the two Ernst potentials \cite{osti_6803745,Neugebauer_1983}. In fact, they have construct $N$-soliton solutions in terms of determinants for rational Ernst potentials on the symmetry axis \cite{PhysRevD.50.6179,PhysRevD.50.4993,Neugebauer_1980,etde_5927960}. However, some of these generating techniques have problems in constructing extreme, sub-extreme and hyper-extreme objects at the same time or even a problem in interpreting the independents' mathematical parameters contained in the solution. One advantage of using the Sibgatullin integral method is the direct relation of the solution parameters with the physical properties of the objects to be described, and, as we shall prove, all parameters that appear to construct the solution have a map one-to-one with the physical parameters, the multipole moments.

Considering axisymmetric stationary spacetimes, the idea of adding solitons to the background is the following. Consider the background as being the Minkowski space. The Ernst  potentials for such spacetime is $\mathcal{E}=1$ and $\Phi=0$. By adding rational functions on the symmetry axis, you are adding solitons into the background. That is, the solution

\begin{equation}
	\mathcal{E}(z,0)=1+\dfrac{e_1}{z-\beta_1},\qquad\Phi(z,0)=\dfrac{f_1}{z-\beta_1}
\end{equation}
is equivalent to a ``1-soliton solution''. Notice that by adding an arbitrary rational function, the solutions are  asymptotically flat (at least on the symmetry axis). The majority of interesting cases fall into the $N$-soliton solutions possessing first order poles. Hence, let us consider the general N-soliton electrovac solution characterized by the Ernst potentials \cite{Ernst} on the symmetry axis given in terms of a polynomial quotient \cite{RMJ}:
\vspace{-0.1cm}
\begin{align}
\mathcal{E}(\rho=0,z)=e(z)=\dfrac{z^N+\sum_{l=1}^{N}{a_lz^{N-l}}}{z^N+\sum_{l=1}^{N}{b_lz^{N-l}}}=\dfrac{P(z)}{R(z)}\label{ePR},\\
\Phi(\rho=0,z)=f(z)=\dfrac{\sum_{l=1}^{N}{c_lz^{N-l}}}{z^N+\sum_{l=1}^{N}{b_lz^{N-l}}}=\dfrac{Q(z)}{R(z)}\label{fPR},
\end{align}

\noi where $a_l,b_l,c_l$, $k=1,\cdots N$ are $3N$ arbitrary complex constants. It is worth noting that the higher order coefficients have been chosen to give an appropriate asymptotic behavior to the potentials at spatial infinity. Moreover, it is assumed that the previous quotients are irreducible and thus $R$ posses only roots of multiplicity one. Or in an equivalent form
\vspace{-0.1cm}
\be e(z)=1+\sum_{l=0}^{N}\dfrac{e_l}{z-\beta_l},\qquad
	f(z)=\sum_{l=1}^{N}\dfrac{f_l}{z-\beta_l},\ee

\noi while the coefficients $a_l,b_l,c_l$ are related to the ones $e_l,f_l,\beta_l$ through the relations

\begin{equation}
\begin{aligned} 
    e_l=\dfrac{P(\beta_l)}{\prod\limits_{k\ne l}^{N}(\beta_l-\beta_k)};\qquad\qquad & f_l=\dfrac{Q(\beta_l)}{\prod\limits_{k\ne l}^{N}(\beta_l-\beta_k)}; \\ 
    R(\beta_l)=0.
\end{aligned}
\end{equation}

Papapetrou has shown that the most general line element for a spacetime with the prescribed symmetries, in the absence of cosmological constant, can be written in the following form \cite{Papapetrou1964zz}:

\begin{equation}\label{PapapetrouLE}
ds^2=F(dt^2-\omega d\varphi)^2-F^{-1}[e^{2\gamma}(d\rho^2+dz^2)+\rho^2d\varphi^2]
\end{equation} 

Here the coordinate system utilized is composed of what is called the Weyl canonical coordinates $(t,\varphi,\rho,z)$. The metric functions depend upon the spatial coordinates  $z\in(-\infty,\infty)$ and $\rho\in[0,\infty)$. The function $\omega$ is related to the angular momentum of the sources, which can represent rotations around the axis at $\rho=0$.

In 1995, E. Ruiz \emph{et al.} \cite{RMJ} found, via the Sibgatullin integral method \cite{Sibgatullin,Manko1993}, a very concise form to write the general solution of this $N$-soliton problem containing $3N$ complex arbitrary parameters. In other words, the Ernst potentials and their corresponding metric function are written in a very simple way. As a matter of fact, this general solution simplifies the study of some particular metrics with a real physically meaningful. The expressions for the metric functions and for the Ernst potentials are given by

\begin{equation}\label{ernstr}
\mathcal{E}(\rho,z)=\dfrac{E_+}{E_{-}};\qquad\qquad \Phi(\rho,z)=\dfrac{W}{E_{-}}
\end{equation}

\begin{equation} E_\pm =
 \begin{vmatrix}
 1 & 1 & \dots & 1\\
 \pm 1 & \frac{r_1}{\alpha_1-\beta_1} & \dots & \frac{r_{2N}}{\alpha_{2N}-\beta_{1}} \\
 \vdots & \vdots & \ddots & \vdots\\
 \pm 1 & \frac{r_1}{\alpha_1-\beta_N} & \dots & \frac{r_{2N}}{\alpha_{2N}-\beta_{N}} \\
 0 & \frac{h_1(\alpha_1)}{\alpha_1-\beta_1^*} & \cdots & \frac{h_1(\alpha_{2N})}{\alpha_{2N}-\beta_1^*}\\
 \vdots & \vdots & \ddots & \vdots\\
 0 & \frac{h_N(\alpha_1)}{\alpha_1-\beta_N^*} & \cdots & \frac{h_N(\alpha_{2N})}{\alpha_{2N}-\beta_N^*}\\
 \end{vmatrix}\qquad\qquad
 W=
 \begin{vmatrix}
 0 & f(\alpha_1) & \dots & f(\alpha_{2N})\\
 -1 & \frac{r_1}{\alpha_1-\beta_1} & \dots & \frac{r_{2N}}{\alpha_{2N}-\beta_{1}} \\
 \vdots & \vdots & \ddots & \vdots\\
 -1 & \frac{r_1}{\alpha_1-\beta_N} & \dots & \frac{r_{2N}}{\alpha_{2N}-\beta_{N}} \\
 0 & \frac{h_1(\alpha_1)}{\alpha_1-\beta_1^*} & \cdots & \frac{h_1(\alpha_{2N})}{\alpha_{2N}-\beta_1^*}\\
 \vdots & \vdots & \ddots & \vdots\\
 0 & \frac{h_N(\alpha_1)}{\alpha_1-\beta_N^*} & \cdots & \frac{h_N(\alpha_{2N})}{\alpha_{2N}-\beta_N^*}\\
 \end{vmatrix}
\end{equation}

\begin{equation}
    F=\frac{D}{2E_{+}E_{-}},
\qquad e^{2\gamma}=\frac{D}{2K_{0}K_{0}^{*}\prod\limits_{n=1}^{2N}r_{n}},
\qquad \omega=\frac{2 \operatorname{Im}\left(E_{-} H^{*}-E^{*}_{-} G-W I^{*}\right)}{D},
\end{equation}
\vspace{-0.1cm}
\begin{align}
D&=E_{+} E^{*}_{-}+E^{*}_{+} E_{-}+2 W W^{*}, \nonu\\\nonu\\
 H&=
 \begin{vmatrix}
 z & 1 & \cdots & 1 \\
 -\beta_{1} & \frac{r_{1}}{\alpha_{1}-\beta_{1}} & \cdots & \frac{r_{2 N}}{\alpha_{2 N}-\beta_{1}} \\
 \vdots & \vdots & \ddots & \vdots \\
 -\beta_{N} & \frac{r_{1}}{\alpha_{1}-\beta_{N}} & \cdots & \frac{r_{2 N}}{\alpha_{2 N}-\beta_{N}} \\
 e^{*}_{1} & \frac{h_{1}\left(\alpha_{1}\right)}{\alpha_{1}-\beta^{*}_{1}} & \cdots & \frac{h_{1}\left(\alpha_{2 N}\right)}{\alpha_{2 N}-\beta^{*}_{1}} \\
 \vdots & \vdots & \ddots & \vdots \\
 e^{*}_{N} & \frac{h_{N}\left(\alpha_{1}\right)}{\alpha_{1}-\beta^{*}_{N}} & \cdots & \frac{h_{N}\left(\alpha_{2 N}\right)}{\alpha_{2 N}-\beta^{*}_{N}}
 \end{vmatrix},\qquad
 G=\begin{vmatrix}
 0 & g_{1} & \dots & g_{2N}\\
 -1 & \frac{r_{1}}{\alpha_{1}-\beta_{1}} & \cdots & \frac{r_{2 N}}{\alpha_{2 N}-\beta_{1}} \\
 \vdots & \vdots & \ddots & \vdots \\
 -1 & \frac{r_{1}}{\alpha_{1}-\beta_{N}} & \cdots & \frac{r_{2 N}}{\alpha_{2 N}-\beta_{N}} \\
 0 & \frac{h_{1}\left(\alpha_{1}\right)}{\alpha_{1}-\beta^{*}_{1}} & \cdots & \frac{h_{1}\left(\alpha_{2 N}\right)}{\alpha_{2 N}-\beta^{*}_{1}} \\
 \vdots & \vdots & \ddots & \vdots \\
 0 & \frac{h_{N}\left(\alpha_{1}\right)}{\alpha_{1}-\beta^{*}_{N}} & \cdots & \frac{h_{N}\left(\alpha_{2 N}\right)}{\alpha_{2 N}-\beta^{*}_{N}}
 \end{vmatrix},\qquad\\
 I&=\begin{vmatrix}
 \sum\limits_{l=1}^{N} f_{l} & 0 & f\left(\alpha_{1}\right) & \dots & f\left(\alpha_{2 N}\right) \\
 z & 1 & 1 & \dots & 1 \\
 -\beta_{1} & -1 & \frac{r_{1}}{\alpha_{1}-\beta_{1}} & \cdots & \frac{r_{2 N}}{\alpha_{2 N}-\beta_{1}} \\
 \vdots & \vdots & \vdots & \ddots & \vdots \\
 -\beta_{N} & -1 & \frac{r_{1}}{\alpha_{1}-\beta_{N}} & \cdots & \frac{r_{2 N}}{\alpha_{2 N}-\beta_{N}} \\
 e^{*}_{1} & 0 & \frac{h_{1}\left(\alpha_{1}\right)}{\alpha_{1}-\beta^{*}_{1}} & \cdots & \frac{h_{1}\left(\alpha_{2 N}\right)}{\alpha_{2 N}-\beta^{*}_{1}} \\
 \vdots & \vdots & \vdots & \ddots & \vdots \\
 e^{*}_{N} & 0 & \frac{h_{N}\left(\alpha_{1}\right)}{\alpha_{1}-\beta^{*}_{N}} & \cdots & \frac{h_{N}\left(\alpha_{2 N}\right)}{\alpha_{2 N}-\beta^{*}_{N}}
 \end{vmatrix}, \qquad
K_{0}=
 \begin{vmatrix}
 1 & \cdots & 1 \\
  \frac{1}{\alpha_{1}-\beta_{1}} & \cdots & \frac{1}{\alpha_{2 N}-\beta_{1}} \\
 \vdots & \vdots & \ddots  \\
  \frac{1}{\alpha_{1}-\beta_{N}} & \cdots & \frac{1}{\alpha_{2 N}-\beta_{N}} \\
  \frac{h_{1}\left(\alpha_{1}\right)}{\alpha_{1}-\beta^{*}_{1}} & \cdots & \frac{h_{1}\left(\alpha_{2 N}\right)}{\alpha_{2 N}-\beta^{*}_{1}} \\
 \vdots & \vdots & \ddots \\
  \frac{h_{N}\left(\alpha_{1}\right)}{\alpha_{1}-\beta^{*}_{N}} & \cdots & \frac{h_{N}\left(\alpha_{2 N}\right)}{\alpha_{2 N}-\beta^{*}_{N}}
 \end{vmatrix}, \\\nonu\\
 g_{n}&=r_{n}+\alpha_{n}-z, \qquad h_{l}(\alpha_{n})=e_{l}^{*}+2f_{l}^{*}f(\alpha_{n}),
\end{align}

\noi where $r_n=\sqrt{\rho^2+(z-\alpha_n)^2}$ are the distances from the value $\alpha_{n}$ defining the location of the sources to any arbitrary point $(\rho,z)$ off the symmetry axis. In this case, $\alpha_n$ are the $2N$ roots satisfying the following characteristic equation:
\be P(z)R^*(z)+P^*(z)R(z)+2Q(z)Q^*(z)=0. \ee

An important point to be underlined from this solution is the fact that at least a priori, these $3N$ complex parameters do not necessarily have a physical meaning unless we link them first to the Geroch-Hansen multipole moments \cite{GerochII,HP,Simon,Sotiriou}. A first development that helped us to contour this problem was a previous work provided by by Hernández-Pastora \cite{pastora} and by Manko \emph{et al.} \cite{MankoRuiz}, in which has been analyzed vacuum solutions ($\Phi=0$) relating the $2N$ parameters $a_l$ and $b_l$ with the corresponding $2N$ multipole moments.

\vspace{-0.1cm}
\section{Relations between the Ernst potentials and multipole moments in vacuum case}\label{multipolevacuum}
\vspace{-0.1cm}
As has been shown in \cite{Simon,HP,fodor2020calculation}, the Geroch-Hansen multipole moments \cite{Hansen,GerochII}, $P_n$ and $Q_n$ , for a given stationary axisymmetric exact solution  can be obtained from their corresponding coefficients expansion of the Ernst potentials $\xi$ and $q$ evaluated on the symmetry axis. That is when $z\rightarrow\infty$, namely,
\vspace{-0.1cm}
\be
\xi=\sum_{k=0}^{\infty}m_{k}z^{-k-1}, \qquad  q=\sum_{k=0}^{\infty}q_{k}z^{-k-1}\label{mqmultipole},
\ee

\noi being $\xi$ and $q$ related to $\mathcal{E}$ and $\Phi$ in the form
\vspace{-0.1cm}
\be \mathcal{E}=\dfrac{1-\xi}{1+\xi}, \qquad \Phi=\dfrac{q}{1+\xi}. \label{ernstxiq} \ee
the coefficients $m_k$ and $q_k$ are related to the multipole moments $P_n$ and $Q_n$. We are interested in showing that indeed it is possible to associate the Geroch-Hansen coefficients, $m_k$ and $q_k$, with the constants $a_l$, $b_l$ and $c_l$, and hence, to characterize physically the Ernst equations. As it is shown in \cite{Hoenselaers86,Simon}, the multipole moments $P_k$ and $Q_k$ can be written in terms of the  the  power  series expansion coefficients $m_k$ and $q_k$, of the Ernst potentials on the symmetry axis\footnote{Once $m_k$ and $q_k$ are known, the multipole moments $P_k$ and $Q_k$ can be constructed or vice-versa.}. The real and imaginary parts of the multipole $P_k$ are associated with the mass and angular multipole moments, respectively \cite{Hansen}. In the same way, the real and imaginary parts of the multipole $Q_k$ are related to the electric and magnetic field multipole moments \cite{Simon}. As pointed out in reference \cite{HP}, it is enough to know the behavior of the Ernst potentials $\mathcal{E}$ and $\Phi$ on the symmetry axis to perform an analytic continuation on them into the whole space \cite{HauserErnst,stephani_kramer_maccallum_hoenselaers_herlt_2003}. Hence, the multipole coefficients $m_k$ and $q_k$ seem to have a significant hole in the present development. Moreover, it is possible to show that the multipole moments uniquely characterize the geometry of the spacetime \cite{Simon-Beig,Simon}. The arbitrariness of these coefficients brings us the question of which might be the condition they should fulfill to satisfy the relation between them and $a_l$, $b_l$ and $c_l$. The substitution of Eqs.\ \eqref{ePR}-\eqref{fPR} into \eqref{mqmultipole} allows us to find
\vspace{-0.1cm}
\bea \dfrac{R(z)-P(z)}{R(z)+P(z)}=\sum_{k=0}^{\infty}m_{k}z^{-k-1},\qquad
\dfrac{2\,Q(z)}{R(z)+P(z)}=\sum_{k=0}^{\infty}q_{k}z^{-k-1}\label{qmutipolerelations}.\eea

Before continuing, first we are going to revisit the analysis employed in references \cite{pastora,MankoRuiz} regarding the vacuum case. So, taking into account only the first equality Eq.\ (\ref{qmutipolerelations}), in the absence of electromagnetic field, after equating the coefficients with the same powers of $z$, we obtain
\vspace{-0.1cm}
\begin{align} \label{mank}
&\frac{1}{2}(b_1-a_1) = m_0, \nonu\\
&\frac{1}{2}(b_2-a_2)=m_1+\frac{1}{2}(b_1+a_1)m_0, \nonu\\
&\vdots\nonu\\
&\frac{1}{2}(b_N-a_N)=m_{N-1}+\frac{1}{2}(b_1+a_1)m_{N-2}+\cdots+\frac{1}{2}(b_{N-1}+a_{N-1})m_0, \nonu\\
&0=m_n+\frac{1}{2}(b_1+a_1)m_{n-1}+\cdots+\frac{1}{2}(b_N+a_N)m_{n-N}\text{, for } n\ge N.
\end{align}

It follows that the simple redefinitions $A_l=\frac{1}{2}(b_l-a_l)$, $B_l=\frac{1}{2}(b_l+a_l)$ with $l=1,2\cdots N$, $A_0=B_0=1$, permits us to get straightforwardly the following system of algebraic equations
\vspace{-0.1cm}
\begin{align}
&A_{n+1}=\sum_{l=0}^{n}B_l\,m_{n-l},\qquad n=0,1,\dots,N-1, \label{Aequantion}\\
&0=\sum_{l=0}^{N}B_l\,m_{n-l},\qquad n\ge N. \label{Bequation}
\end{align}

The above set of equations constitute an algebraic system with infinite equations for a finite number of variables, which may give us the opportunity to write $a_l$ and $b_l$ in terms of $m_n$. In order to describe the $N$-soliton problem, the authors in \cite{RMJ} used a set of $2N$ arbitrary parameters, $a_l$ and $b_l$. Thus, in principle, it would be possible to use a set of 2$N$ coefficients $m_k$ to describe such a problem. In what follows, it is outlined a generalization of the compatibility condition which ensures a similar system.

In order to find the $N$ variables $B_l$, $N$ equations are needed. Then, inside the infinity set of parameters $m_n$, take $N$ elements $\{m_{n_1},m_{n_2},\dots,m_{n_N}\}$ in such way that $n_i\ge N$ with $i=1,2,\dots,N$. Consider, now, the equation \eqref{Bequation} for this set of $n_i$.
\vspace{-0.1cm}
\be\sum_{l=1}^{N}B_l\,m_{n_{i}-l}=-m_{n_{i}}\label{Bequation2},\ee

\noi or equivalently

\vspace{-0.1cm}
\be
\begin{pmatrix}
    m_{n_{1}-1} & m_{n_{1}-2} & \cdots & m_{n_{1}-N+1} & m_{n_{1}-N}\\
    m_{n_{2}-1} & m_{n_{2}-2} & \cdots & m_{n_{2}-N+1} & m_{n_{2}-N}\\
    \vdots      & \vdots      & \ddots & \vdots        & \vdots\\
    m_{n_{N}-1} & m_{n_{N}-2} & \cdots & m_{n_{N}-N+1} & m_{n_{N}-N}\\
    \end{pmatrix}
    \begin{pmatrix}
    B_1\\
    B_2\\
    \vdots\\
    B_N
    \end{pmatrix}
    =-
    \begin{pmatrix}
    m_{n_{1}}\\
    m_{n_{2}}\\
    \vdots\\
    m_{n_{N}}\\
    \end{pmatrix}\label{Bequation3},
    \ee

Defining a new object, $L_i$, as a $i\times i$ matrix:

\vspace{-0.1cm}
\be
L_i=\begin{pmatrix}
m_{n_{1}-N+i-1} & m_{n_{2}-N+i-1} & \cdots & m_{n_{i-1}-N+i-1} & m_{n_{i}-N+i-1}\\
m_{n_{1}-N+i-2} & m_{n_{2}-N+i-2} & \cdots & m_{n_{i-1}-N+i-2} & m_{n_{i}-N+i-2}\\
\vdots      & \vdots      & \ddots & \vdots        & \vdots\\
m_{n_{1}-N} & m_{n_{2}-N} & \cdots & m_{n_{i-1}-N} & m_{n_{i}-N}\\
\end{pmatrix},
\ee

 Where it is straightforward to observe that the system \eqref{Bequation3} only have solution when $\det L_N\ne 0$. In order to shortening the notation on the following equations, let us use $|-|$ to correspond the determinant of a matrix.  Using the Cramer's rule to find the coefficients $B_l$, we obtain:

\vspace{-0.1cm}
\be
B_l=(-1)^l |L_N|^{-1}
\begin{vmatrix}
m_{n_{1}} & m_{n_{1}-1} & \cdots & m_{n_{1}-(l-1)} & m_{n_{1}-(l+1)} & \cdots & m_{n_{1}-N}\\
m_{n_{2}} & m_{n_{2}-1} & \cdots & m_{n_{2}-(l-1)} & m_{n_{2}-(l+1)} & \cdots & m_{n_{2}-N}\\
\vdots      & \vdots      & \ddots & \vdots        & \vdots & \vdots\\
m_{n_{N}} & m_{n_{N}-1} & \cdots & m_{n_{N}-(l-1)} & m_{n_{N}-(l+1)} & \cdots & m_{n_{N}-N}\\
\end{vmatrix}.
\ee

Due to this high symmetric structure, it is possible to rewrite $B_l$ as:

\vspace{-0.1cm}
\be
B_l=|L_N|^{-1}
\begin{vmatrix}
0 & m_{n_{1}} & m_{n_{2}} & \cdots & m_{n_{N-1}} & m_{n_{N}}\\
0 &           &           &        &             &\\
\vdots &           &           &        &             &\\
1 &           &           &  L_N      &             & \\
\vdots &           &           &        &             &\\
0 &           &           &        &             & \\
\end{vmatrix},
\ee

\noi where the row corresponding to the ``1'' in the first column  is the ($l+1$)-th row. By using this result for $B_l$ the coefficients $A_l$ can be found by means of Eq.\ \eqref{Aequantion} and , in a similar way, can be written as:

\be
A_{l+1}=|L_N|^{-1}
\begin{vmatrix}
m_l & m_{n_{1}} & m_{n_{2}} & \cdots & m_{n_{N-1}} & m_{n_{N}}\\
m_{l-1} &           &           &        &             &\\
\vdots &           &           &        &             &\\
m_0 &           &           &  L_N      &             & \\
\vdots &           &           &        &             &\\
0 &           &           &        &             & \\
\end{vmatrix}.
\ee

Nonetheless, $A_l$ and $B_l$ are described in terms of, at most, $N^2+2N$ independent coefficients $m_k$ and by hypothesis it asserts that they must be written in terms of $2N$  coefficients $m_k$. Therefore, it is necessary to restrict the set $\{m_{n_{i}}\}$, which is solution of the equation \eqref{Bequation2}, and its condition is only respected in case when the set $\{m_{n_{i}}\}$ is chosen  with $n_1=N$, $n_2=N+1$, $\dots$, $n_N=2N-1$.  Then, $|L_i|$ can be written as: 

\vspace{-0.1cm}
\be
|L_i|=\begin{vmatrix}
m_{i-1} & m_{i} & \cdots & m_{2i-3} & m_{2i-2}\\
m_{i-2} & m_{i-1} & \cdots & m_{2i-4} & m_{2i-3}\\
\vdots      & \vdots      & \ddots & \vdots        & \vdots\\
m_{0} & m_{1} & \cdots & m_{i-2} & m_{i-1}\\
\end{vmatrix},
\ee

 And then, $A_l$ and $B_l$, written in terms of $2N$ $m_k$'s, take the form:
 
\vspace{-0.1cm}
\be
A_{l+1}=|L_N|^{-1}
\begin{vmatrix}
m_l & m_{N} & m_{N+1} & \cdots & m_{2N-2} & m_{2N-1}\\
m_{l-1} &           &           &        &             &\\
\vdots &           &           &        &             &\\
m_0 &           &           &  L_N      &             & \\
\vdots &           &           &        &             &\\
0 &           &           &        &             & \\
\end{vmatrix},
\ee

\be
B_l=|L_N|^{-1}
\begin{vmatrix}
0 & m_{N} & m_{N+1} & \cdots & m_{2N-2} & m_{2N-1}\\
0 &           &           &        &             &\\
\vdots &           &           &        &             &\\
1 &           &           &  L_N      &             & \\
\vdots &           &           &        &             &\\
0 &           &           &        &             & \\
\end{vmatrix}.
\ee

Notice that in $B_l$ the only elements that do not repeat are $m_0$ and $m_{2N-1}$, since the diagonals on the principal direction are constituted by equal elements, with the exception of the elements in the first column. Because of that, $m_1$ appears twice, $m_2$ appears three times, until $m_{N-1}$ and $m_N$ which appear $N$ times, and then, $m_{N+1}$ appears $N-1$ times and so on.

For completeness, we are going to deduce now the condition that coefficients must satisfy when calculating $L_i$ for $i>N$. In order to complete this statement, let us now consider
\vspace{-0.1cm}
\be
|L_{N+1}|=\begin{vmatrix}
m_{N} & m_{N+1} & \cdots & m_{2N-1} & m_{2N}\\
m_{N-1} & m_{N} & \cdots & m_{2N-2} & m_{2N-1}\\
\vdots      & \vdots      & \ddots & \vdots        & \vdots\\
m_{0} & m_{1} & \cdots & m_{N-1} & m_{N}\\
\end{vmatrix},
\ee

\noi and after making a cofactor expansion, we find:

\vspace{-0.1cm}
\be
|L_{N+1}|=(-1)^N| L_N| \sum_{k=0}^{N}B_k m_{2N-k}.
\ee

However, by hypothesis, the sum $\sum\limits_{k=0}^{N}B_k m_{n-k}$ is equal to zero for all $n\ge N$. Thus, $m_k$ must be in such way that $|L_{N+1}|=0$. By induction, it is straightforward to see that all determinants $|L_n|=0$ for all $n> N$, proving the following lemma, which was first stated  in  reference \cite{pastora} and revisited in \cite{MankoRuiz} reads:

\begin{lemma} \label{lemma1}
    Given all multipole coefficients $m_i$, once fixed the set of coefficients $\{m_{i}\}$ with $i=0,1,\dots, 2N-1$, the necessary and sufficient condition for this set to describe the behaviour of the Ernst potentials on the symmetry  axis as a polynomial quotient \eqref{ePR} is that the determinant $|L_n|$ be nonzero for $n=N$ and zero for all $n>N$.
\end{lemma}

\vspace{-0.2cm}
\section{Relations between the Ernst potentials and multipole moments in electrovacuum case}\label{multipoleelectrovacuum}
\vspace{-0.2cm}
As mentioned before,  Lemma \ref{lemma1} is not a new result, however, its general proof might remained unnoticed in literature since it was first presented in the thesis \cite{pastora} but it became well-known in \cite{MankoRuiz} without proof. As a matter of fact, one may bear in mind to consider a similar analysis outlined in the vacuum case but now applied to the electrovacuum case, which to our knowledge has been ignored due to its complexity, in order to extend the result already given in \cite{pastora}. In order to try to generalize these results for the cases where the electromagnetic field is present, a similar analysis for the equation \eqref{qmutipolerelations} will be done, so that in the end the $3N$ variables $a_l$, $b_l$ and $c_l$ can be written in terms of $3N$ coefficients $m_k$ and $q_k$ related with the multipole moments. Equating the coefficients with the same powers of $z$, we find:
\vspace{-0.1cm}
\begin{align}
&c_1 = q_0,\nonu\\
&c_2=q_1+\frac{1}{2}(b_1+a_1)q_0\nonu\\
&\vdots\nonu\\
&c_N=q_{N-1}+\frac{1}{2}(b_1+a_1)q_{N-2}+\cdots+\frac{1}{2}(b_{N-1}+a_{N-1})q_0\nonu\\
&0=q_n+\frac{1}{2}(b_1+a_1)q_{n-1}+\cdots+\frac{1}{2}(b_N+a_N)q_{n-N}, \quad \text{, for } n\ge N.
\end{align}

Such system can be summarized into:
\vspace{-0.1cm}
\begin{align}
&c_{n+1}=\sum_{l=0}^{n}B_l\,q_{n-l},\qquad n=0,1,\dots,N-1,\label{cequantion}\\
&0=\sum_{l=0}^{N}B_l\,q_{n-l},\qquad n\ge N. \label{Bqequation}
\end{align}

Notice that the above Eqs.\ (\ref{cequantion})-(\ref{Bqequation}) have the same structure as the Eqs.\ \eqref{Aequantion}-\eqref{Bequation}. In addition, the same function $B_l$, which was already evaluated in terms of the coefficients $m_k$, will be evaluated now in terms of the coefficients $q_k$. Since it is the same function $B_l$, when it is written in terms of  $m_k$ or $q_k$ it must be equivalent.

Due to the fact that equation's structure contains the same aspect as in the vacuum case, we will introduce a new index here to the determinant $L_i$ in order to differentiate whether it is written in terms of $m_k$ or $q_k$, that is, $L_{i,m}$ and $L_{i,q}$. That is:

\begin{equation}
L_{i,q}=\begin{pmatrix}
q_{i-1} & q_{i} & \cdots & q_{2i-3} & q_{2i-2}\\
q_{i-2} & q_{i-1} & \cdots & q_{2i-4} & q_{2i-3}\\
\vdots      & \vdots      & \ddots & \vdots        & \vdots\\
q_{0} & q_{1} & \cdots & q_{i-2} & q_{i-1}\\
\end{pmatrix}.
\end{equation}

 Therefore, the equation for $B_l$ in terms of $q_k$ is given by:
 
\vspace{-0.1cm}
\be
B_l=|L_{N,q}|^{-1}
\begin{vmatrix}
0 & q_{N} & q_{N+1} & \cdots & q_{2N-2} & q_{2N-1}\\
0 &           &           &        &             &\\
\vdots &           &           &        &             &\\
1 &           &           &  L_{N,q}      &             & \\
\vdots &           &           &        &             &\\
0 &           &           &        &             & \\
\end{vmatrix}.
\ee

Since the variable $B_l$ must be the same independently of whether it is written in terms of the $m_k$ or $q_k$, the following relation is obtained:

\begin{align}\label{Bcondition}
&B_l=\\&|L_{N,m}|^{-1}
\begin{vmatrix}
0 & m_{N} & m_{N+1} & \cdots  & m_{2N-1}\\
0 &           &           &                &\\
\vdots &           &           &                   &\\
1 &           &           &  L_{N,m}                 & \\
\vdots &           &           &                  &\\
0 &           &           &                 & \\
\end{vmatrix}
=|L_{N,q}|^{-1}
\begin{vmatrix}
0 & q_{N} & q_{N+1} & \cdots &  q_{2N-1}\\
0 &           &           &                 &\\
\vdots &           &           &                 &\\
1 &           &           &  L_{N,q}                & \\
\vdots &           &           &                 &\\
0 &           &           &               & \\
\end{vmatrix}\nonumber
\end{align}

Given that the value of $B_l$ is defined the for $m_k$ coefficients, from the equality above, we conclude that, for a fixed $l$, the set of variables $q_k$, loose one degree of freedom. Furthermore, knowing that $l$ ranges from 1 to $N$, one notices that $N$ of the $|q_k|$ variables, for $k=0,1\cdots 2N-1$, are not free, i.e., $N$ variables from the set $q_k$ can be described as a function of $2N$ variables from $m_k$ and $N$ variables of $q_k$. Consequently, one can generalize lemma \ref{lemma1}.

\begin{lemma}\label{lemma2}
   Given all multipole coefficients $m_i$ and $q_i$, once fixed a set of coefficients $\{m_{i}\}$ with $i=0,1,\dots, 2N-1$ and a subset of $N$ coefficients $q_i$ contained in $\{q_i\}$ with $i=0,1,\dots, 2N-1$, the necessary and sufficient conditions for those $3N$ variables to describe the behaviour of the Ernst potentials on the symmetry axis as polynomial quotient \eqref{ePR} and \eqref{fPR}  is that the determinant $|L_n,_{m}^{q}|$ be nonzero for $n=N$ and zero for all $n>N$ and that equation \eqref{Bcondition} is valid.
\end{lemma}

The final representation for the $3N$ variables $A_l$, $B_l$ e $c_l$ written in terms of $2N$ coefficients $m_k$ and $N$ coefficients $q_k$ is given by
\vspace{-0.1cm}
\be
A_{l+1}=|L_{N,m}|^{-1}
\begin{vmatrix}\label{Al}
m_l & m_{N} & m_{N+1} & \cdots & m_{2N-2} & m_{2N-1}\\
m_{l-1} &           &           &        &             &\\
\vdots &           &           &        &             &\\
m_0 &           &           &  L_{N,m}      &             & \\
\vdots &           &           &        &             &\\
0 &           &           &        &             & \\
\end{vmatrix},
\ee

\vspace{-0.1cm}
\be\label{Bl}
B_l=|L_{N,m}|^{-1}
\begin{vmatrix}
0 & m_{N} & m_{N+1} & \cdots & m_{2N-2} & m_{2N-1}\\
0 &           &           &        &             &\\
\vdots &           &           &        &             &\\
1 &           &           &  L_{N,m}      &             & \\
\vdots &           &           &        &             &\\
0 &           &           &        &             & \\
\end{vmatrix},
\ee

\vspace{-0.1cm}
\be\label{cl}
c_l=|L_{N,m}|^{-1}
\begin{vmatrix}
q_l & m_{N} & m_{N+1} & \cdots & m_{2N-2} & m_{2N-1}\\
q_{l-1} &           &           &        &             &\\
\vdots &           &           &        &             &\\
q_0 &           &           &  L_{N,m}      &             & \\
\vdots &           &           &        &             &\\
0 &           &           &        &             & \\
\end{vmatrix}.
\ee

\noi where the relation between the variables $A_l$ e $B_l$ with the variables $a_l$ e $b_l$ defined in \eqref{ePR} and \eqref{fPR} is given by
\be\label{ab}
a_l=B_l-A_l, \qquad b_l=B_l+A_l.
\ee

Thereby, we conclude the Section showing that the electrovacuum $N$-soliton solution on the z-axis can be written in terms of the multipole moments in the following compact way:
\be
P(z)=z^{N}+\sum_{l=1}^{N}a_{l}z^{N-l} = z^{N}+\sum_{l=1}^{N}(B_{l}-A_{l})z^{N-l}=
\ee

\be
	=\sum_{l=0}^{N}B_{l}z^{N-l}-\sum_{l=1}^{N}\sum_{k=0}^{l-1}B_{k}m_{l-1-k}z^{N-l} = \sum_{l=0}^{N}B_{l}z^{N-l}-\sum_{l=0}^{N-1}\sum_{k=0}^{l}B_{k}m_{l-k}z^{N-l-1}
\ee
\be
		=\sum_{l=0}^{N}B_{l}z^{N-l}-\sum_{l=0}^{N-1}B_l\sum_{k=0}^{N-1-l}m_{k}z^{N-l-k-1},
\ee

where the first and second terms from the above equality can be written as
\be
\sum_{l=0}^{N}B_{l}z^{N-l}=|L_N|^{-1}
\begin{vmatrix}
z^N & m_{N} & m_{N+1} & \cdots & m_{2N-2} & m_{2N-1}\\
z^{N-1} &           &           &        &             &\\
\vdots &           &           &   L_N     &             &\\
z &           &           &       &             & \\
1 &           &           &        &             & \\
\end{vmatrix},
\ee

\be
\sum_{l=0}^{N-1}B_l\sum_{k=0}^{N-1-l}m_{k}z^{N-l-k-1} =|L_{N,m}|^{-1}
\begin{vmatrix}
\sum\limits_{k=0}^{N-1}m_{k}z^{N-k-1} & m_{N} & m_{N+1} & \cdots & m_{2N-1}\\[10pt]
\sum\limits_{k=0}^{N-2}m_{k}z^{N-k-2} &           &                &             &\\
\vdots &           &           &   L_{N,m}                 &\\
m_0 &           &           &                 & \\
0 &           &           &                 & \\
\end{vmatrix},
\ee

Hence, we can write $P(z)$, $R(z)$ and $Q(z)$ in the very simple form
\be
P(z)=|L_{N,m}|^{-1}
\begin{vmatrix}
z^{N}-\sum\limits_{k=0}^{N-1}m_{k}z^{N-k-1} & m_{N} & m_{N+1} & \cdots & m_{2N-2} & m_{2N-1}\\[10pt]
z^{N-1}-\sum\limits_{k=0}^{N-2}m_{k}z^{N-k-2} &           &           &        &             &\\
\vdots &           &           &   L_{N,m}     &             &\\
z-m_0 &           &           &       &             & \\
1 &           &           &        &             & \\
\end{vmatrix},
\ee

\be
R(z)=|L_{N,m}|^{-1}
\begin{vmatrix}
z^{N}+\sum\limits_{k=0}^{N-1}m_{k}z^{N-k-1} & m_{N} & m_{N+1} & \cdots & m_{2N-2} & m_{2N-1}\\[10pt]
z^{N-1}+\sum\limits_{k=0}^{N-2}m_{k}z^{N-k-2} &           &           &        &             &\\
\vdots &           &           &   L_{N,m}     &             &\\
z+m_0 &           &           &       &             & \\
1 &           &           &        &             & \\
\end{vmatrix},
\ee

\be
Q(z) =|L_{N,m}|^{-1}
\begin{vmatrix}
\sum\limits_{k=0}^{N-1}q_{k}z^{N-k-1} & m_{N} & m_{N+1} & \cdots & m_{2N-2} & m_{2N-1}\\[10pt]
\sum\limits_{k=0}^{N-2}q_{k}z^{N-k-2} &           &           &        &             &\\
\vdots &           &           &   L_{N,m}     &             &\\
q_0 &           &           &       &             & \\
0 &           &           &        &             & \\
\end{vmatrix}.
\ee

Possessing in hands Lemma \ref{lemma1} and \ref{lemma2}, in order to ensure that the multipole coefficients will describe a solution of the Ernst potentials which are rational on the symmetry axis, we need all coefficients $m_i$ and $q_i$, since we need to ensure that  $|L_N,_{m}^{q}|$ be nonzero for $n=N$ and zero for all $n>N$. However, the solution will only be described in terms of $3N$ variables. This lead us to the next part of this work that is to prove that, given a set of $3N$ multipole coefficients, we they will describe a behaviour of a Ernst potentials which are rational on the symmetry axis.

\subsection{Multipole moments of the N-Soliton solution}

So far, we have given the relations and conditions for writing the $3N$ parameters of the N-soliton solution, $a_l$, $b_l$ and $c_l$, in terms of the multipole coefficients $m_l$ and $q_l$. Now, a stronger result can be achieved by studying the inverse relation of these coefficients. That is, we will show that , in fact, $L_{N+1},_{m}^{q}$ is always zero for the N-soliton solution, and the conditions in Lemmas \ref{lemma1} and  \ref{lemma2} are always satisfied for such solution. For this purpose, consider the series below:

\begin{equation}
	\dfrac{\sum\limits_{l=1}^{N}e_l z^{N-l}}{z^N+\sum\limits_{k=1}^{N}d_k z^{N-k}}
\end{equation}

This series has the same shape as in equation \eqref{qmutipolerelations} (they are the same apart from a factor). Therefore, in order to write $m_l$ and $q_l$ in terms of $a_l$, $b_l$ and $c_l$, it is necessary to see how to expand the above series in terms of negative powers of $z$. By canceling the term $z^N$ and focusing on the denominator, we notice that it is possible to expand it in the following power series

\begin{equation}
	\dfrac{1}{1+\sum\limits_{k=1}^{N}d_k z^{-k}}=\sum_{j=0}^{\infty}(-1)^j\left(\sum_{k=1}^{N}d_k z^{-k}\right)^j.
\end{equation}

However

\begin{equation}
	(d_1 z^{-1}+d_2 z^{-2}+\cdots+ d_N z^{-N})^j=\sum_{k_1+k_2+\cdots+k_N=j} \dfrac{j!}{k_1!k_2!\cdots k_N!}\prod_{t=1}^{N}(d_t z^{-t})^{k_t},
\end{equation}

therefore

\begin{equation}
	\dfrac{\sum\limits_{l=1}^{N}e_l z^{N-l}}{z^N+\sum\limits_{k=1}^{N}d_k z^{N-k}}=\sum\limits_{l=1}^{N}e_l z^{-l}\sum_{j=0}^{\infty}\sum_{k_1+k_2+\cdots+k_N=j} \dfrac{j!}{k_1!k_2!\cdots k_N!}\prod_{t=1}^{N}(d_t z^{-t})^{k_t}.
\end{equation}

Now, we need to find the general coefficient for this power series. That is, we must write

\begin{equation}
	\dfrac{\sum\limits_{l=1}^{N}e_l z^{N-l}}{z^z+\sum\limits_{k=1}^{N}d_k z^{N-k}}\equiv \sum_{\alpha=0}^{\infty}h_\alpha z^{-\alpha-1},
\end{equation}

and find the coefficients $h_\alpha$. After some simple calculations, we find

\begin{equation}
	h_{\alpha}=\sum_{l=1}^{N}e_l \theta_{l,\alpha},
\end{equation}

where

\begin{equation}
	\theta_{l,\alpha}=\left\{\begin{matrix}
		&0,  \mbox{ if } \alpha < l,\\
		&\sum\limits_{k_1+2 k_2+\cdots+N k_N=\alpha-l}(-1)^{k_1+k_2+\cdots+k_N}\dfrac{(k_1+k_2+\cdots+k_N)!}{k_1!k_2!\cdots k_N!}\prod\limits_{t=1}^{N}(d_t)^{k_t}, \mbox{ if }  \alpha\ge l.
	\end{matrix}\right.
\end{equation}

From the above equation, it is possible to find a relation between the $h_\alpha$

\begin{equation}
	h_{\alpha+N}=-\sum_{l=1}^{N}d_l h_{\alpha+N-l}.
\end{equation}

This shows that $h_{\alpha+N}$ is a linear combination of the set $\{h_\alpha,h_{\alpha+1},\cdots, h_{\alpha+N-1}\}$ with fixed coefficients $d_l$. This implies that the last column of the matrix whose determinant is $L_{N+1},_{m}^{q}$ is a linear combination of the first $N$ columns. Finally, we can write:

\begin{equation}\label{multipolecoe}
	m_{\alpha}=\frac{1}{2}\sum_{l=1}^{N}(b_l-a_l) \theta_{l,\alpha}, \qquad\qquad q_{\alpha}=\sum_{l=1}^{N}c_l \theta_{l,\alpha},
\end{equation}

\begin{equation}
\theta_{l,\alpha}=\left\{\begin{matrix}
		&0,  \mbox{ if } \alpha < l,\\
		&\sum\limits_{k_1+2 k_2+\cdots+N k_N=\alpha-l}(-1)^{k_1+k_2+\cdots+k_N}\dfrac{(k_1+k_2+\cdots+k_N)!}{k_1!k_2!\cdots k_N!}\prod\limits_{t=1}^{N}\left(\dfrac{b_l+a_l}{2}\right)^{k_t}, \mbox{ if }  \alpha\ge l.
	\end{matrix}\right.
\end{equation}

For this reason, not only $|L_{N+1},_{m}^{q}|$ but $L_{N+k},_{m}^{q}\ k\geq 1,$ is zero for all $N$-soliton solutions. Moreover, this implies that all $m_n$ are determined for $n\ge 2N$, and $q_n$ are determined for $n\ge N$. Finally, using this, we can improve Lemma \ref{lemma2} and state the best form of our result.

\begin{theorem}\label{theo1}
	As in Lemma \ref{lemma2} fix a set of coefficients $\{m_{i}\}_{i=1}^{2N}$ and a subset $\{q_{i_1},\ q_{i_2},\ \dots , q_{i_N}\}\subset\{q_i\}_{i=1}^{2N}$. Then, these $3N$ variables describe the behavior of Ernst potentials on the symmetry axis as polynomial quotient \eqref{ePR} and \eqref{fPR}  if and only if the determinant $|L_N,_{m}^{q}|\neq0$.
\end{theorem}

Another interesting outcome evolves from these previous results. First, notice that  Lemma \ref{lemma2}  imposes two conditions in the multipoles in order to describe a $N$-soliton solution, the equation \eqref{Bcondition} be valid and $|L_N,_{m}^{q}|\neq0$. However, we proved that equation \eqref{Bcondition} is always valid in the case of a $N$-soliton solution. Arriving in the Theorem \ref{theo1}, which states that $|L_N,_{m}^{q}|\neq0$ is the only condition to establish the relation between the multipoles and the solution. But notice that, if we have $|L_N,_{m}^{q}|=0$ and $|L_{N-1},_{m}^{q}|\neq0$, we are describing a system of $N-1$ soliton, and the equations are still valid. However, due to the results in the present Section, if a given stationary axisymmetric spacetime, solution of the Einstein-Maxwell equations, such that its multipole moments satisfy \eqref{Bcondition} and $|L_N,_{m}^{q}|\neq0$, then this solution can be approximated as a $N$-soliton solution. Moreover, it worth mentioned that it is simple to construct exact spacetimes possessing finite coefficients $m_k$ and $q_k$. When $a_l=-b_l$ for $l=1,2,\cdots, N$, then the value of $\theta_{l,\alpha}$ is $0$ for $l\neq\alpha$ and $1$ for $l=\alpha$.

In order to exemplify how the results of the present paper should be interpreted and used in practical terms, we give a series of examples below.

\section{Solutions from prescribed multipole moments}\label{examples}
\subsection{2-Soliton solution}

Based on the $2$-soliton solution, let us find the solution associated with the given multipoles moments. In order to clarify how the method derived in the present paper should be used, we will start with know multipole moments are written and then construct the metric solution associated with them. The $2$-soliton solution can describe several interesting cases, from the Tomimatsu-Sato with $\delta=2$ to two interacting Kerr-Newman-like black holes \cite{TomiSato,1981GReGr..13..195N,KRAMER1980259,Herdeiro_2008,CABRERAMUNGUIA2020135945}. However, the physical parameterization of these solutions is somehow obscure due to the fact that the direct relation between the multipole coefficients and the $N$-soliton solutions has not been made until now.   Consider then, the Ernst potentials of the $2$-soliton solution in the symmetry axis:

\begin{align}
	\mathcal{E}(\rho=0,z)=\dfrac{z^2+a_1 z+a_2}{z^2+b_1 z+b_2},\\
	\Phi(\rho=0,z)=\dfrac{c_1 z+c_2}{z^2+b_1 z+b_2}
\end{align}

Notice here that the $2$-soliton solution is given in terms of $6$ arbitrary parameters $\{a_l, b_l, c_l\}$, $l=1,2$. Let us see how these parameters are connected to the multipole moments. Using the equation \eqref{multipolecoe}, the multipole coefficients are written as:

\begin{align*}
	&m_0=\dfrac{b_1-a_1}{2}\,,\\
	&m_1=\frac{a_1^2-b_1^2+2 b_2-2 a_2}{4}\,,\\
	&m_2=\frac{b_1^3-a_1^3-a_1^2 b_1+a_1 \left(4 a_2+b_1^2\right)-4 b_1 b_2}{8}\,,\\
	&m_3=\frac{ (a_1^2-b_1^2)  \left((a_1+b_1)^2-4 (a_2+b_2)\right)-2 (a_2-b_2) \left((a_1+b_1)^2-2 (a_2+b_2)\right)}{16}\,,
\end{align*}

And

\begin{align*}
	&q_0=c_1\,,\\
	&q_1=c_2-\frac{c_1 (a_1+b_1)}{2}\,,\\
	&q_2=\frac{c_1 (a_1+b_1)^2-2 c_2 (a_1+b_1)-2 c_1 (a_2+b_2)}{4}\,,\\
	&q_3=\frac{4 c_1 (a_1+b_1) (a_2+b_2)-c_1 (a_1+b_1)^3+2 c_2 (a_1+b_1)^2-4 c_2 (a_2+b_2)}{8} \,.
\end{align*}

Theorem \ref{theo1} give us the necessary and sufficient conditions for the coefficients $m_k's$ and $q_k's$ describe a $2$-soliton solution. With their form in hands (written in terms of $\{a_l, b_l, c_l\}$), it is easy to show from the above discussion that it is needed only 6 parameters to describe the Ernst potentials. First of all, notice that:

\begin{equation}
	L_{n,m}=L_{n,q}=0\,, \quad \forall\,n \ge 3.
\end{equation}

and

\begin{equation}
	L_{2,m}\neq0,\qquad L_{2,q}\neq0.
\end{equation}

\begin{equation}
	(L_{2,m})^{-1}
	\begin{vmatrix}
		0 & m_{2} & m_{3} \\
		1 & m_{1} & m_{2} \\
		0 & m_{0} & m_{1} \\
	\end{vmatrix}
	=(L_{2,q})^{-1}
	\begin{vmatrix}
		0 & q_{2} & q_{3} \\
		1 & q_{1} & q_{2} \\
		0 & q_{0} & q_{1} \\
	\end{vmatrix}
\end{equation}

\begin{equation}
	(L_{2,m})^{-1}
	\begin{vmatrix}
		0 & m_{2} & m_{3} \\
		0 & m_{1} & m_{2} \\
		1 & m_{0} & m_{1} \\
	\end{vmatrix}
	=(L_{2,q})^{-1}
	\begin{vmatrix}
		0 & q_{2} & q_{3} \\
		0 & q_{1} & q_{2} \\
		1 & q_{0} & q_{1} \\
	\end{vmatrix}
\end{equation}

Hence, the above parameterization in fact satisfies the Theorem \ref{theo1}; and therefore describe a $2$-soliton solution.  And trough equations \eqref{Al}, \eqref{Bl} and \eqref{cl}, we can find the inverse relation:

\begin{align*}
	&a_1=\frac{-m_{2} m_{0}^2+m_{1}^2 m_{0}-m_{3} m_{0}+m_{1} m_{2}}{m_{0} m_{2}-m_{1}^2}\,,\qquad a_2=\frac{m_{1}^3+(m_{3}-2 m_{0} m_{2}) m_{1}-m_{2}^2+m_{0}^2 m_{3}}{m_{0} m_{2}-m_{1}^2}\,,\\
	&b_1=\frac{m_{1}^3+(m_{3}-2 m_{0} m_{2}) m_{1}-m_{2}^2+m_{0}^2 m_{3}}{m_{0} m_{2}-m_{1}^2}\,,\qquad b_2=\frac{m_{1}^3 + m_{2}^2 + m_{0}^2 m_{3} -
		m_{1} (2 m_{0} m_{2} + m_{3})}{m_{1}^2 - m_{0} m_{2}}\,.
\end{align*}

Finally, the coefficients $c_l$ can be found by \eqref{cl}:

\begin{equation}
	c_1=q_0\,,\qquad c_2=\frac{m_{1}^2 q_{1}-m_{1} m_{2} q_{0}+m_{0} (m_{3} q_{0}-m_{2} q_{1})}{m_{1}^2-m_{0} m_{2}}\,.
\end{equation}

\subsubsection{Tomimatsu-Sato solution with \texorpdfstring{$\delta$=$2$}{d=2}}

Now, let us proceed in the same way for the particular case of multipoles moments associated with the Tomimatsu-Sato solution \cite{TomiSato}. We will find that the Ernst potentials on the symmetry axis are recovered from the corresponding multipole moments.  In order to clarify how the method derived in the preceding sections should be used, consider the multipole moments below and then construct the metric solution associated with them. Due to the complexity of higher orders, we will consider a solutions with only the first 6 multipoles $P_n$:

\begin{align*}
	&P_0=M\,,\\
	&P_1=i a M\,,\\
	&P_2=-\frac{1}{4} M \left(3 a^2+M^2\right)\,,\\
	&P_3=-\frac{1}{2} i a M \left(a^2+M^2\right)\,,\\
	&P_4=\frac{1}{112} M \left(35 J^4+66 J^2 M^2+11 M^4\right)\,,\\
	&P_5=\frac{1}{112} M \left(35 J^4+66 J^2 M^2+11 M^4\right)\,.
\end{align*}

Therefore, the coefficients $m_k$ are:

\begin{align*}
	&m_0=M\,,\\
	&m_1=ia M\,,\\
	&m_2=-\frac{1}{4} M \left(3 a^2+M^2\right)\,,\\
	&m_3=-\frac{1}{2} i a M \left(a^2+M^2\right)\,,\\
	&m_4=\frac{1}{16} \left(5 a^4 M+10 a^2 M^3+M^5\right)\,,\\
	&m_5=\frac{1}{16} i \left(3 a^5 M+10 a^3 M^3+3 a M^5\right)\,.
\end{align*}

With the coefficients $m_k$ in hand, it is easy to show from the discussion above that only 4 parameters are needed to describe the Ernst potentials. First of all, notice that:

\begin{equation}
	L_{n,m}=0\,, \quad \forall\,n \ge 3.
\end{equation}

and

\begin{equation}
	L_{2,m}=\frac{1}{4} \left(M^4-a^2 M^2\right)\,.
\end{equation}

In order to relate the multipole coefficients $m_k$ and $q_k$ with the Ernst coeficients $a_l$ and $b_l$ the following relations must be true:

\begin{equation}
	B_1=(L_{2,m})^{-1}
	\begin{vmatrix}
		0 & m_{2} & m_{3} \\
		1 & m_{1} & m_{2} \\
		0 & m_{0} & m_{1} \\
	\end{vmatrix}=-i a
\end{equation}

\begin{equation}
	B_2=(L_{2,m})^{-1}
	\begin{vmatrix}
		0 & m_{2} & m_{3} \\
		0 & m_{1} & m_{2} \\
		1 & m_{0} & m_{1} \\
	\end{vmatrix}=\frac{1}{4} \left(M^2-a^2\right)
\end{equation}

which can be verified after a straightforward calculation. Consequently:

\begin{align*}
	&B_1=-i a\,,\qquad B_2=\frac{1}{4} \left(M^2-a^2\right)\,,\\
	&A_1=M\,,\qquad A_2=0\,.
\end{align*}

Using the equation \eqref{ab} yields:

\begin{align*}
	&a_1=- (M+ia)\,,\qquad a_2=\frac{1}{4} \left(M^2-a^2\right)\,,\\
	&b_1=M-i a\,,\qquad b_2=\frac{1}{4} \left(M^2-a^2\right)\,.
\end{align*}

Thus, the Ernst potentials $\mathcal{E}$  can now be evaluated with the relations \eqref{ab}:

\begin{equation}
	\mathcal{E}=\dfrac{1-\xi}{1+\xi}=\dfrac{z^2-(M+i a)z+\frac{M^2-a^2}{4}}{z^2+(M-i a)z+\frac{M^2-a^2}{4}}
\end{equation}
recovering, then, the Ernst potential on the symmetry axis  for the Tomimatsu-Sato solution with distorsion parameter $\delta=2$ \cite{TomiSato}.

\subsection{Approximated solutions}

We already saw how the multipole moments, under certain conditions, are exactly matched to the $N$-soliton solution. We will elucidate how to find approximated solutions describing physical objects possessing the required multipole moments by applying the methodology derived in the present paper can be utilized to construct physical objects with the required multipole moments or, at least, find approximated solutions. Hence, consider the first seven gravitational multipole moments as given in \cite{fodor2020calculation}:

\begin{align}
 P_0&=m_0 \ , \label{eqp0m0} \\
 P_1&=m_1 \ , \\
 P_2&=m_2 \ , \\
 P_3&=m_3+\frac{1}{5}q^{*}_0 S_{10} \ , \\
 P_4&=m_4
 -\frac{1}{7}m^{*}_0 M_{20}
 +\frac{3}{35}q^{*}_1 S_{10}
 +\frac{1}{7}q^{*}_0(3S_{20}-2H_{20}) \ , \\
 P_5&=m_5
 -\frac{1}{21}m^{*}_1 M_{20}
 -\frac{1}{3}m^{*}_0 M_{30}
 +\frac{1}{21}q^{*}_2 S_{10}
 +\frac{1}{21}q^{*}_1(4S_{20}-3H_{20}) \notag\\
 &+\frac{1}{21}q^{*}_0\left(
 q^{*}_0 q_0 S_{10}
 -m^{*}_0 m_0 S_{10}
 +14S_{30}+13S_{21}-7H_{30}
 \right) \ , \\
 P_6&=m_6
 -\frac{5}{231}m^{*}_2 M_{20}
 -\frac{4}{33}m^{*}_1 M_{30}
 +\frac{1}{33}m^{*2}_0 m_0 M_{20}
 -\frac{1}{33}m^{*}_0(18M_{40}+8M_{31}) \notag\\
 &+\frac{1}{33}q^{*}_3 S_{10}
 +\frac{1}{231}q^{*}_2(25S_{20}-20H_{20})
 +\frac{2}{231}q^{*}_1(35S_{30}+37S_{21}-21H_{30}) \notag\\
 &-\frac{1}{1155}(37q^{*}_1m^{*}_0
  +13q^{*}_0m^{*}_1)m_0 S_{10}
 +\frac{1}{33}q^{*2}_0
  \left(5q_0 S_{20}-4m_0 Q_{20}+3q_1 S_{10}\right) \notag\\
 &+\frac{10}{231}q^{*}_1q^{*}_0 q_0 S_{10}
 +\frac{2}{33}q^{*}_0m^{*}_0
  \left(2m_0 H_{20}-3q_0 M_{20}-2m_1 S_{10}\right) \\
 &+\frac{1}{33}q^{*}_0
  \left(30S_{40}+32S_{31}-24H_{31}-12H_{40}\right) \ . \notag
\end{align}
And the electromagnetic moments:
\begin{align}
 Q_0&=q_0 \ , \\
 Q_1&=q_1 \ , \\
 Q_2&=q_2 \ , \\
 Q_3&=q_3-\frac{1}{5}m^{*}_0 H_{10} \ , \\
 Q_4&=q_4
 +\frac{1}{7}q^{*}_0 Q_{20}
 -\frac{3}{35}m^{*}_1 H_{10}
 -\frac{1}{7}m^{*}_0(3H_{20}-2S_{20}) \ , \\
 Q_5&=q_5
 +\frac{1}{21}q^{*}_1 Q_{20}
 +\frac{1}{3}q^{*}_0 Q_{30}
 -\frac{1}{21}m^{*}_2 H_{10}
 -\frac{1}{21}m^{*}_1(4H_{20}-3S_{20}) \notag\\
 &+\frac{1}{21}m^{*}_0\left(
 m^{*}_0 m_0 H_{10}
 -q^{*}_0 q_0 H_{10}
 -14H_{30}-13H_{21}+7S_{30}
 \right) \ , \\
 Q_6&=q_6
 +\frac{5}{231}q^{*}_2 Q_{20}
 +\frac{4}{33}q^{*}_1 Q_{30}
 +\frac{1}{33}q^{*2}_0 q_0 Q_{20}
 +\frac{1}{33}q^{*}_0(18Q_{40}+8Q_{31}) \notag\\
 &-\frac{1}{33}m^{*}_3 H_{10}
 -\frac{1}{231}m^{*}_2(25H_{20}-20S_{20})
 -\frac{2}{231}m^{*}_1(35H_{30}+37H_{21}-21S_{30})\notag\\
 &-\frac{1}{1155}(37m^{*}_1q^{*}_0
  +13m^{*}_0q^{*}_1)q_0 H_{10}
 +\frac{1}{33}m^{*2}_0
  \left(5m_0 H_{20}-4q_0 M_{20}+3m_1 H_{10}\right) \notag\\
 &+\frac{10}{231}m^{*}_1m^{*}_0 m_0 H_{10}
 +\frac{2}{33}m^{*}_0q^{*}_0
  \left(2q_0 S_{20}-3m_0 Q_{20}-2q_1 H_{10}\right) \label{eqq6q6}\\
 &-\frac{1}{33}m^{*}_0
  \left(30H_{40}+32H_{31}-24S_{31}-12S_{40}\right) \ . \notag
\end{align}

Where

\begin{align}
 M_{ij}&=m_i m_j-m_{i-1}m_{j+1} \ , &
 S_{ij}&=m_i q_j-m_{i-1}q_{j+1} \ , \label{eqmijsij}\\
 Q_{ij}&=q_i q_j-q_{i-1}q_{j+1} \ , &
 H_{ij}&=q_i m_j-q_{i-1}m_{j+1} \ , \label{eqqijhij}
\end{align}

Notice here that the Multipole moment $P_{n}$ ($Q_{n}$) is linear on $m_{n}$ ($q_{n}$) and do not depend on the higher orders of the coefficients $m_{n}$ ($q_{n}$). Hence, we can chose  $m_{n}$ ($q_{n}$) so that the multipole moment $P_{n}$ ($Q_{n}$)  is the desired one. If, for instance, we want to describe a pole-dipole source, that is, a source only possessing a mass and electric
monopole moment and angular momentum and magnetic dipole moments, we can set all $m_k$ and $q_k$ such that \cite{HP}:

\begin{equation}
\begin{aligned}
  &P_0=m,\qquad \qquad Q_0=e,\\
 &P_1=i m a,\qquad\qquad Q_1=i e \mu,\\
 &P_n=0,\qquad\qquad Q_n=0, \quad \text{for all }n\ge 2.
\end{aligned}
\end{equation}

By means of the general equations \eqref{eqp0m0}-\eqref{eqq6q6}, we can determine the first seven multipole coefficients $m_k$ and $q_k$ univocally:

\begin{equation}\label{mdipole}
\begin{aligned}
  &m_0=m,\quad m_1=i m a,\quad m_2=0,\quad m_3=-\frac{1}{5} i e^2 m (a-\mu ),\\
 &m_4=\dfrac{1}{7}a^2m^3-\dfrac{8}{35}a m e^2 \mu+\dfrac{3}{35} m e^2\mu^2,\\
 &m_5=\frac{3}{35} i m e^2 (a-\mu ) (e^2-m^2)-\dfrac{1}{21}i a m(a^2 m^2-e^2\mu^2),\\
 &m_6=\frac{1}{21} a^2 m^3 \left(m^2-2 e^2\right)-\frac{2}{35}e^2 \mu  m^3 (a-\mu )+\frac{1}{105} e^4 \mu  m (16 a-11 \mu ),
\end{aligned}
\end{equation}

and

\begin{equation}\label{qdipole}
\begin{aligned}
  &q_0=e,\quad e_1=i e \mu,\quad q_2=0,\quad q_3=-\frac{1}{5} i e m^2 (a-\mu ),\\
 &q_4=-\dfrac{1}{7}\mu^2e^3+\dfrac{8}{35}a m^2 e \mu-\dfrac{3}{35}e a^2 m^2,\\
 &q_5=\frac{3}{35} i e m^2 (a-\mu ) (e^2-m^2)+\frac{1}{21} i e \mu  (e^2 \mu^2 -a^2m^2),\\
 &q_6=\frac{1}{21} e^3 \mu ^2 \left(e^2-2 m^2\right)+\frac{2}{35} a e^3 m^2 (a-\mu )+\frac{1}{105} a e m^4 (16 \mu -11 a).
\end{aligned}
\end{equation}

Hence, although the multipole moments of order higher than 1 are zero, the multipole coefficients are not. Notice that the coefficients presented here differ from those presented on the papers \cite{HP, Sotiriou} based on the recent paper of Fodor et al \cite{fodor2020calculation}. It is interesting to compare our solution with the previous one in the limit $\mu=a$:

\begin{equation}
\begin{aligned}
  &m_0=m,\quad m_1=i m a,\quad m_2=0,\quad m_3=0,\\
 &m_4=\frac{1}{7} a^2 m \left(m^2-e^2\right),\\
 &m_5=-\frac{1}{21} i a^3 m \left(m^2-e^2\right),\\
 &m_6=\frac{1}{21} a^2 m \left(e^2-m^2\right)^2,
\end{aligned}
\end{equation}

and

\begin{equation}
\begin{aligned}
  &q_0=e,\quad e_1=i e \mu,\quad q_2=0,\quad q_3=0,\\
 &q_4=-\frac{1}{7} a^2 e \left(e^2-m^2\right),\\
 &q_5=\frac{1}{21} i a^3 e \left(e^2-m^2\right),\\
 &q_6=\frac{1}{21} a^2 e \left(e^2-m^2\right)^2.
\end{aligned}
\end{equation}

This agrees with the previous results presented in \cite{HP,Sotiriou} (although the authors only presented the coefficients until the fifth-order). But, if, instead of considering the limit $\mu=a$, we consider $e^2=m^2$, the coefficients do not agree. Under the special case of $\mu=a$ and $e^2=m^2$, all coefficients $m_k$ and $q_k$ vanish for $k\ge2$. For the special case, these multipole coefficients describe a $N$-soliton solution. In fact, the expansion 

\begin{equation}
    \xi(\rho=0,z)=\dfrac{m}{z}+i \dfrac{a m}{z^2},
\end{equation}
\begin{equation}
    q(\rho=0,z)=\pm\dfrac{m}{z}\pm i \dfrac{a m}{z^2},
\end{equation}
implies

\begin{equation}
    \mathcal{E}(\rho=0,z)=\dfrac{z^2-m z-i a m}{z^2+m z+ i a m},
\end{equation}
\begin{equation}
    \Phi(\rho=0,z)=\dfrac{\pm m z\pm i a m}{z^2+m z+ i a m},
\end{equation}

Which describes a $2$-soliton solution, and the solution for the whole spacetime is given by equation \eqref{ernstr}. 

However, for the general case where $e^2\ne m^2$ and $\mu\ne a$, it does not seem that the determinants $L_N,_{m}^{q}$ become zero for some order $N$. Therefore, the solution for the monopole-dipole source can not be exactly described as a $N$-soliton solution. Yet, we can use the $N$-soliton solution in order to approach the monopole-dipole solution. We can successfully describe the monopole-dipole solution as a $N$-soliton solution up to the order $N$. For instance, the corresponding $7$-soliton solution based on formulae \eqref{multipolecoe} is simply given by

\begin{equation}
    \mathcal{E}(\rho=0,z)=\dfrac{z^7-m_0 z^6-m_1 z^5-m_2 z^4-m^3 z^3-m_4 z^2-m_5 z-m_6}{z^7+m_0 z^6+m_1 z^5+m_2 z^4+m^3 z^3+m_4 z^2+m_5 z+m_6}
\end{equation}
\begin{equation}
    \Phi(\rho=0,z)=\dfrac{q_0 z^6+q_1 z^5+q_2 z^4+q^3 z^3+q_4 z^2+q_5 z+q_6}{z^7+m_0 z^6+m_1 z^5+m_2 z^4+m^3 z^3+m_4 z^2+m_5 z+m_6}
\end{equation}
where $m_k$ and $q_k$ are given by equations \eqref{mdipole} and \eqref{qdipole}. And again, we can construct the solution in the whole space by means of equation \eqref{ernstr}. This solution will have

\begin{equation}
\begin{aligned}
  &P_0=m,\qquad \qquad Q_0=e,\\
 &P_1=i m a,\qquad\qquad Q_1=i e \mu,\\
 &P_n=0,\qquad\qquad Q_n=0, \quad \text{for }6\ge n\ge 2,\\
  &P_n\neq0,\qquad\qquad Q_n\neq0, \quad \text{for } n\ge 6.
\end{aligned}
\end{equation}

That is, the above identification allows us to describe an approximate solution for the monopole-dipole source in charged spacetimes. For the vacuum case, see for instance reference \cite{HernandezPastora:1998mc}

\section{Conclusions}

Solutions of the Einstein-Maxwell equations for stationary axisymmetric spacetimes proved to be not only interesting from the theoretical point of view but also to have experimental applications. Keeping in mind that, we briefly revised the family of solutions introduced by Manko and Ruiz \cite{RMJ} named ``Extended N-soliton solution'' focusing on clarifying what is understood as soliton in the context of general Relativity.

The present paper was devoted to constructing a direct relation between the multipole moments and the mathematical parameters which appear.  We also demonstrated how to construct exact solutions from their given multipole moments in the general $N$-soliton case. That is, this work extends the previous development of Manko and Ruiz to include electromagnetic fields. As mentioned by the authors in \cite{MankoRuiz}, ``It is remarkable that (4.4) is a linear system of algebraic equations when either one looks for the form of $m_n$ in terms of $a_l$ and $b_l$ , or vice versa, when one wants to see how the constants $a_l$ and $b_l$ depend on $m_n$'' (equation 4.4 in their paper is equivalent to equation \eqref{mank}). We could extend their sentence to include the parameters $q_n$ and $c_l$ and also say that it is remarkable that the physical parameterization of the $N$-soliton solution relies on solving a linear system.

 In this way, the direct link was made between the coefficients of the multipole expansion in General Relativity and the $3N$ parameters of the $N$ -soliton solution. This result has been summarized in the Lemmas of Sections \ref{multipolevacuum} and \ref{multipoleelectrovacuum}, and are extensions of the Lemma already presented in the literature by the above-mentioned authors. Furthermore, the theorem presented in the same Section shows that any set of multipole moments satisfying a not so restrict condition can build an $N$-soliton solution.
 
 Notice that for particular cases, we could have a solution having, for instance, $|L_n,_{m}|$ be nonzero for $n=N_1$ and zero for all $n>N_1$; and $|L_n,_{q}|$ be nonzero for $n=N_2$ and zero for all $n>N_2$ for $N_1\neq N_2$. All results remains valid and we would describe a $N_2$-soliton solution if, for instance, $N_2> N_1$.
 
 Although compact, the results of the present paper can get even simple when symmetries are considered. Consider, for instance, spacetimes possessing equatorial symmetry/antisymmetry. Pachón-Contreras et al. \cite{Pach_n_2006} and Ernst et al. \cite{Ernst_2006,Ernst_2007} deduced how such symmetries are expressed in terms of the Ernst potentials on the symmetry axis. They found that, for the symmetric case, $m_k$ must be real for $k$ even and imaginary for $k$ odd. While $q_k$ has its behavior depending on a parameter $\epsilon$. When $\epsilon=1$, $q_k$ obeys the same rule: it must be real for $k$ even and imaginary for $k$ odd.  On the other hand, for $\epsilon=-1$, $q_k$ must be imaginary for $k$ even and real for $k$ odd. For the antisymmetric case, $m_k$ vanishes for $k$ odd and $q_k$ vanishes for $k$ even. It is straightforward to see that equations \eqref{Al}-\eqref{cl}, where the mathematical parameters of the solution, $a_l$, $b_l$ and $c_l$, are written in terms of the determinants of the physical parameters, the multipole coefficients, $m_k$ and $q_k$, become even simpler! In fact, the proofs contained in reference \cite{Ernst_2006} become quite more direct if one makes use of equation \eqref{multipolecoe}, deduced in the present paper.
 
 We also conclude that a generic solution of Einstein’s equation coupled with electromagnetism, whose multipole moments satisfy this condition, can be approximated as an $N$-soliton solution.

\ack
One of the authors, ESCF , would like to thanks Professors Betti Hartmann and Gyula Fodor for their support and helpful comments during the preparation of the present work. This study was financed in part by the Coordenação de Aperfeiçoamento de Pessoal de Nível Superior - Brasil (CAPES)-Finance Code 001. ICM acknowledges the financial support of SNI-CONACyT, M\'exico, grant with CVU No. 173252.

\section*{References}

\bibliographystyle{hhieeetr}
\bibliography{multi_soliton}

\end{document}